\documentclass[11pt]{article}
\usepackage{amsmath}
\usepackage{amsfonts}
\usepackage[english]{babel}
\usepackage{amssymb}
\usepackage{graphicx}
\usepackage[a4paper]{geometry}
\usepackage{hyperref}

\newcommand{\be}{\begin{equation}}
\newcommand{\ee}{\end{equation}}
\newcommand{\bea}{\begin{eqnarray}}
\newcommand{\eea}{\end{eqnarray}}
\newcommand{\vsp}{\vspace{0.4cm}}

\newcommand{\lag}{\mathfrak{L}}

\newtheorem{remark}{Remark}

\title{Hamilton-Jacobi approach to Potential Functions in Information Geometry}
\author{Florio M. Ciaglia, Fabio Di Cosmo\\ 
\textit{Dipartimento di Fisica, Universit\`a di Napoli ``Federico II"}\\
\textit{Via Cinthia Edificio 6, I-80126 Napoli, Italy}\\
\textit{and INFN-Sezione di Napoli, Via Cinthia Edificio 6, I-80126 Napoli, Italy}\\ 
\and Domenico Felice, Stefano Mancini\\
\textit{School of Science and Technology, University of Camerino}\\
\textit{Via Madonna delle Carceri 7, I-62032 Camerino, Italy}\\
\textit{and INFN-Sezione di Perugia, Via A. Pascoli, I-06123 Perugia, Italy}\\ 
\and Giuseppe Marmo\\ 
\textit{Dipartimento di Fisica, Universit\`a di Napoli ``Federico II"}\\
\textit{Via Cinthia Edificio 6, I-80126 Napoli, Italy}\\
\textit{and INFN-Sezione di Napoli, Via Cinthia Edificio 6, I-80126 Napoli, Italy}\\ 
\and Juan M. P\'{e}rez-Pardo\\
\textit{INFN-Sezione di Napoli, Via Cinthia Edificio 6, I-80126 Napoli, Italy}\\
\textit{and Departamento de Matem\'{a}ticas, Universitad Carlos III de Madrid,}\\
\textit{Avda. de la Universitad 30, Legan\'{e}s, Madrid, Spain}
}
\date{}

\begin{document}
\maketitle

\abstract{
The search for a potential function $S$ allowing to reconstruct a given metric tensor $g$ and a given symmetric covariant tensor $T$ on a manifold $\mathcal{M}$ is formulated as the Hamilton-Jacobi problem associated with a canonically defined Lagrangian on $T\mathcal{M}$.
The connection between this problem, the geometric structure of the space of pure states of quantum mechanics, and the theory of contrast functions of classical information geometry is outlined.
}

\section{Introduction}

The recent development of quantum information theory has led to a growing interest in the geometrical description of the space of quantum states (\cite{bengtsson_zyczkowski-geometry_of_quantum_states:_an_introduction_to_quantum_entanglement}).
In this direction, the geometrical approach to quantum mechanics developed for example in \cite{ashtekar_schilling-geometrical_formulation_of_quantum_mechanics,ercolessi_marmo_morandi-from_the_equations_of_motion_to_the_canonical_commutation_relations}, allows to reformulate the information encoded in the scalar product $\langle\,,\rangle$ of the Hilbert space $\mathcal{H}$ of the system in terms of two tensors on the space of pure states $\mathcal{P}(\mathcal{H})$, namely, a metric tensor $g$ called the Fubini-Study metric, and a symplectic form $\omega$.
The manifold $\mathcal{P}(\mathcal{H})$ together with these tensors forms what is known as a K\"{a}hler manifold, i.e., $\mathcal{P}(\mathcal{H})$ admits a complex structure $J$ such that the metric tensor $g$ and the symplectic form $\omega$ are mutually related according to the following compatibility condition

\be
g\left(X\,,J(Y)\right)=\omega\left(X\,,Y\right)\,,
\ee
where $X$ and $Y$ are arbitrary vector fields on $\mathcal{P}(\mathcal{H})$.
These two tensors together define a Hermitian tensor $h=g + \mathrm{i}\omega$ on $\mathcal{P}(\mathcal{H})$ (\cite{ashtekar_schilling-geometrical_formulation_of_quantum_mechanics, ercolessi_marmo_morandi-from_the_equations_of_motion_to_the_canonical_commutation_relations, provost_vallee-riemannian_structure_on_manifolds_of_quantum_states}).
The geometric information of a K\"{a}hler manifold is completely encoded in the so-called (local) K\"{a}hler potential $K$.
This is a (local) function on the manifold which allows to recover the explicit expression of $h$, and thus of $g$ and $\omega$, as follows:

\be
h_{j\bar{k}}=\frac{\partial^{2} K}{\partial z^{j}\partial \bar{z}^{k}}\,,
\ee
where $z^{j},\bar{z}^{k}$ are holomorphic coordinates on the K\"{a}hler manifold.
As it is clear, the way in which $h$ is extracted from the K\"{a}hler potential $K$ highly depends on the complex structure $J$ on $\mathcal{P}(\mathcal{H})$ by means of the derivatives with respect to the holomorphic coordinates.

Following \cite{ercolessi_marmo_morandi-from_the_equations_of_motion_to_the_canonical_commutation_relations}, the pullback of $h$ to the Hilbert space $\mathcal{H}$ reads:

\be
\widetilde{h}=\pi^{*}h=\frac{\langle\mathrm{d}\psi|\mathrm{d}\psi\rangle}{\langle\psi|\psi\rangle} -\frac{\langle\mathrm{d}\psi|\psi\rangle\langle\psi|\mathrm{d}\psi\rangle}{\langle\psi|\psi\rangle^{2}}\,,
\ee
where the projection map $\pi\colon \mathcal{H}\rightarrow\mathcal{P}(\mathcal{H})$ is given by:

\be
\pi\left(|\psi\rangle\right)=\frac{|\psi\rangle\langle\psi|}{\langle\psi|\psi\rangle}\,.
\ee
To get a more concrete feeling of $\widetilde{h}$, let us consider a finite-dimensional Hilbert space $\mathcal{H}\cong\mathbb{C}^{n}$ and write a vector in $\mathcal{H}$ as $|\psi\rangle=\sqrt{p_{j}}\,\mathrm{e}^{\mathrm{i}\varphi_{j}}\,|e_{j}\rangle$, where $(p_{1},...,p_{n})$ is a probability vector, that is, $\sum\,p_{j}=1$, $\mathrm{e}^{\mathrm{i}\varphi_{j}}$ is a phase factor, and summation on $j$ is understood.
Then we have:

$$
\widetilde{h}=\frac{1}{4}\left[\langle \mathrm{d}(\ln \vec{p})\otimes \mathrm{d}(\ln \vec{p})\rangle_{\vec{p}} - \langle \mathrm{d}(\ln \vec{p})\rangle_{\vec{p}}\otimes \langle\mathrm{d}(\ln \vec{p})\rangle_{\vec{p}} \right] +
$$
\be
+ \langle \mathrm{d}\vec{\varphi} \otimes\mathrm{d}\vec{\varphi}\rangle_{\vec{p}} - \langle \mathrm{d}\vec{\varphi}\rangle \otimes \langle \mathrm{d}\vec{\varphi}\rangle_{\vec{p}} + \frac{\mathrm{i}}{2}\left[\langle \mathrm{d} \left(\ln \vec{p}\right)\wedge\mathrm{d}\vec{\varphi}\rangle_{\vec{p}} - \langle \mathrm{d}\left(\ln \vec{p}\right)\rangle_{\vec{p}}\wedge\langle\mathrm{d}\vec{\varphi}\rangle_{\vec{p}}\right]\,,
\ee
where $\langle\,\rangle_{\vec{p}}$ denotes the expectation value with respect to the probability vector $\vec{p}$.
The real part of this tensor is symmetric, and defines the pullback to $\mathcal{H}$ of the Fubini-Study metric $g$, while the imaginary part is antisymmetric and defines the pullback to $\mathcal{H}$ of the symplectic form $\omega$.
Note that (the pullback of) $g$ is composed of two terms, the first one is equivalent to the Fisher-Rao metric on the space of probability vectors $(p_{1}\,...\,p_{n})$, while the second term can be interpreted as a quantum contribution to the Fisher-Rao metric due to the phase of the state.
In \cite{facchi_kulkarni_manko_marmo_sudarshan_ventriglia-classical_and_quantum_fisher_information_in_the_geometrical_formulation_of_quantum_mechanics} it is shown that, once a particular submanifold of pure states $\mathcal{M}\subseteq \mathcal{P}(\mathcal{H})$  is chosen, the Hermitian tensor $h$ induces a tensor $h_{\mathcal{M}}$ on $\mathcal{M}$, which can be degenerate.
The real part of $h_{\mathcal{M}}$ defines a metric tensor $g_{\mathcal{M}}$ which, again, can be interpreted as a quantum analog of the Fisher-Rao metric on $\mathcal{M}$.
However, $\mathcal{M}$ is not necessarily a K\"{a}hler submanifold of the space of pure states, hence, the existence of a K\"{a}hler potential there is not guaranteed.

Since the idea of using a potential function to describe geometrical structures naturally fits into the conceptual framework of the geometrical formulation of quantum mechanics, it makes sense to ask wether there is a ``potential-like function'' for the quantum Fisher-Rao metric on $\mathcal{M}$. Because we cannot rely on complex coordinates, a possible potential function should be expressed in terms of real coordinates on the real manifold $\mathcal{M}$.
Interestingly, a similar problem is deeply investigated in the geometrical approach to classical information theory pioneered by Amari \cite{amari_nagaoka-methods_of_information_geometry}, where the Fisher-Rao metric of a statistical manifold is recovered by means of a so-called contrast function $S$.
This is a two-point function, i.e., a function on $\mathcal{M}\times\mathcal{M}$, that, analogously to the K\"{a}hler potential $K$, contains all the geometrical information of the statistical manifold.
Furthermore, in \cite{zhang_li-symplectic-and_kahler_structures_on_statistical_manifolds_induced_from_divergence_functions}, the possibility of using the contrast function $S$ of information geometry in order to define a K\"{a}hler structure on $\mathcal{M}\times\mathcal{M}$ is analyzed.

It is then natural to look at this classical case with a more ``quantum'' attitude in order to unveil differences and analogies that could lead to a bidirectional flow of ideas and mathematical tools between the two settings. For instance, since there is no complex structure $J$ in the classical setting, we think that a more thorough investigation of this situation could help to understand and point out what the role of $J$ in the quantum setting is. 
Such a programme is highly non-trivial, and cannot be accomplished in a single work.

In this contribution we want to analyze the problem of finding a potential function for a statistical manifold in the framework of classical information geometry from a different point of view.
We will formulate the problem in a geometric and dynamical framework common to both the classical and quantum setting, and we will show that a solution of the Hamilton-Jacobi problem for a canonically-defined Lagrangian $\lag$ is actually a potential function for a given statistical manifold $\mathcal{M}$.
We point out that this formulation of the problem does not depend on the fact that $\mathcal{M}$ is a statistical manifold in the sense of information geometry, that is, our formulation can be naturally applied to cases in which the metric tensor on $\mathcal{M}$ is not the classical Fisher-Rao metric.
The results presented here suggest that it could be possible to think of well-known divergence functions, such as the Kullback-Leibler divergence, as the Hamilton principal function of some suitable Lagrangian.

\section{Hamilton-Jacobi theory and potential functions}\label{sec: H-J theory and potential functions}

Following Lauritzen (\cite{a.a.v.v.-differential_geometry_in_statistical_inference}) a statistical manifold is a triple $(\mathcal{M},\,  g  , T)$, where $\mathcal{M}$ is a differential manifold whose points parametrize a family of probability distributions, $g$ is a metric tensor on $\mathcal{M}$, and $T$ is a symmetric covariant tensor of order $3$ on $\mathcal{M}$, called the skewness tensor. 
Starting with the tensor $T$ and the Christoffel symbols $_{g}\Gamma$ of the Levi-Civita connection of $g$, it is possible to define a family of affine torsionless connections $\nabla^{\alpha}$ whose Christoffel symbols are:

\be
_{\alpha}\Gamma_{jkl}:=\,_{g}\Gamma_{jkl} - \frac{\alpha}{2}T_{jkl}\,,
\label{alpha-connections}
\ee
where $T_{jkl}$ and $_{g}\Gamma_{jkl}$ are, respectively, the components of $T$ and $_{g}\Gamma$.

In the context of classical information geometry, we have a measure space $\mathcal{X}$, the space $\mathcal{P}(\mathcal{X})$ of probability distributions on $\mathcal{X}$, and a statistical manifold $\mathcal{M}$ is identified with a submanifold of $\mathcal{P}(\mathcal{X})$ by means of an injective map:

\be
\mathcal{M}\ni\mathrm{m}\mapsto \mathrm{p}(\mathbf{x}\,,\mathrm{m})\,\mathrm{d}x\in\mathcal{P}(\mathcal{X})\,,
\ee
where $\mathrm{d}x$ is a reference measure on $\mathcal{X}$.
Then, introducing a coordinate system $\{\xi^{j}\}$ on $\mathcal{M}$, the Fisher-Rao metric $g$, and the skewness tensor $T$ can be obtained by:

\be\label{eqn: fisher-rao metric}
g_{jk}=\int_{\mathcal{X}}\,\mathit{p}(\mathbf{x}\,,\xi)\,\left(\frac{\partial\log(\mathit{p})}{\partial\xi^{j}}\right)\left(\frac{\partial\log(\mathit{p})}{\partial\xi^{k}}\right)\,\mathrm{d}x\,,
\ee

\be\label{eqn: canonical classical skewness tensor}
T_{jkl}=\int_{\mathcal{X}}\,\mathit{p}(\mathbf{x}\,,\xi)\,\left(\frac{\partial\log(\mathit{p})}{\partial\xi^{j}}\right)\left(\frac{\partial\log(\mathit{p})}{\partial\xi^{k}}\right)\left(\frac{\partial\log(\mathit{p})}{\partial\xi^{l}}\right)\,\mathrm{d}x\,.
\ee

\vsp

For every pair $(\alpha\,,-\alpha)$, the connections $\nabla^{\alpha}$ and $\nabla^{-\alpha}$  are dual with respect to the metric $g$ in the sense that the following equation holds:

\be
Z\left(g(X\,,Y)\right)=g\left(\nabla^{\alpha}_{Z}\,X\,,Y\right) + g\left(X\,,\nabla^{-\alpha}_{Z}\,Y\right)
\ee
for all vector fields $X\,,Y\,,Z$ on $\mathcal{M}$.
The Levi-Civita connection $\nabla_{g}=\nabla^{0}$ is the only self-dual torsionless connection with respect to the metric $g$, and a statistical manifold for which $T=0$ is called self-dual.

\begin{remark}\label{first remark}
A statistical structure on $\mathcal{M}$ can be alternatively defined using $g$ and the pair of torsionless dual connections $\nabla\equiv\nabla^{1},\nabla^{*}\equiv\nabla^{-1}$ (\cite{amari_nagaoka-methods_of_information_geometry, a.a.v.v.-differential_geometry_in_statistical_inference}).
In this framework, if the connections $\nabla$ and $\nabla^{*}$ are flat connections, the statistical manifold $(\mathcal{M}\,,g\,,\nabla\,,\nabla^{*})$ is called dually flat.
This obviously poses some topological obstructions on $\mathcal{M}$, for instance, it turns out that the space of pure states of a finite-level quantum system does not admit a dually flat structure (see \cite{ay_tuschmann-dually_flat_manifolds_and_global_information_geometry, ay_tuschmann-duality_versus_flatness_in_quantum_information_geometry}).
According to equation \eqref{alpha-connections}, it is clear that the statistical manifolds $(\mathcal{M}\,,g\,,T)$ and $(\mathcal{M}\,,g\,,\nabla\,,\nabla^{*})$ are completely equivalent.
\end{remark}

It is possible to prove \cite{amari_nagaoka-methods_of_information_geometry, a.a.v.v.-differential_geometry_in_statistical_inference, matumoto-any_statistical_manifold_has_a_contrast_function} that the geometrical structure of every statistical manifold can be completely encoded in a two-point function $S\colon \mathcal{M}\times \mathcal{M}\rightarrow\mathbb{R}$ called contrast function.
This is a distance-like function such that:

\be\label{defining condition 1} 
S(m_{1}\,,m_{2})\geq0 \;\;\;\;\forall \; m_{1},m_{2}\,,
\ee
\be\label{defining condition 2} 
S(m_{1}\,,m_{2})=0\;\;\;\; \mbox{{\itshape iff }}m_{1}=m_{2}\,.
\ee
Here, the first $\mathcal{M}$ is thought of as the manifold of initial points whose coordinates we denote by $q_{\mathrm{in}}$, and the second $\mathcal{M}$ is the manifold of final points whose coordinates we denote by $q_{\mathrm{fin}}$.
If $S$ is at least $C^{3}$, \cite{matumoto-any_statistical_manifold_has_a_contrast_function}, it follows that:

\be\label{eqn: sufficient condition for second derivative to form a tensor}
\left.\frac{\partial S}{\partial q^{j}_{\mathrm{in}}}\right|_{q_{\mathrm{in}}=q_{\mathrm{fin}}}=\left.\frac{\partial S}{\partial q^{j}_{\mathrm{fin}}}\right|_{q_{\mathrm{in}}=q_{\mathrm{fin}}}=0\,.
\ee
The metric $g$ and the tensor $T$ are recovered from it as follows:

\be
\left.\frac{\partial^{2}\,S}{\partial q^{j}_{\mathrm{in}}\partial q^{k}_{\mathrm{in}}}\right|_{q_{\mathrm{in}}=q_{\mathrm{fin}}}=\left.\frac{\partial^{2}\,S}{\partial q^{j}_{\mathrm{fin}}\partial q^{k}_{\mathrm{fin}}}\right|_{q_{\mathrm{in}}=q_{\mathrm{fin}}}=-\left.\frac{\partial^{2}\,S}{\partial q^{j}_{\mathrm{fin}}\partial q^{k}_{\mathrm{in}}}\right|_{q_{\mathrm{in}}=q_{\mathrm{fin}}}=g_{jk}\,,
\label{metric}
\ee

\be
\left. \dfrac{\partial ^3 S}{\partial q^{l}_{\mathrm{in}} \partial q^{k}_{\mathrm{fin}} \partial q^{j}_{\mathrm{fin}} }\right|_{q_{\mathrm{in}}=q_{\mathrm{fin}}} - \left. \dfrac{\partial ^3 S}{\partial q^{l}_{\mathrm{fin}} \partial q^{k}_{\mathrm{in}} \partial q^{j}_{\mathrm{in}} }\right|_{q_{\mathrm{in}}=q_{\mathrm{fin}}} = T_{jkl}\,.
\label{tensor}
\ee
Note that the restriction of the second and third derivatives of $S$ to the diagonal define  tensor fields because of equation \eqref{eqn: sufficient condition for second derivative to form a tensor}.
It is important to note that  $S$ is never unique, and this leads to the need for the definition of a contrast function which is canonical in some suitable sense \cite{ay_amari-a_novel_approach_to_canonical_divergences_within_information_geometry}.

What we propose is to interpret the task of finding a canonical potential function for the statistical manifold $(\mathcal{M}\,,g\,,T)$ in the context of Hamilton-Jacobi theory associated with a particular Lagrangian built directly from the metric $g$ and the symmetric tensor $T$.
Note that the dynamical approach to potential functions presented here is purely geometric in the sense that it relies only on the geometrical structure of $\mathcal{M}$.
This means that $\mathcal{M}$ needs not to be a statistical manifold endowed with the Fisher-Rao metric and the canonical skewness tensor of information geometry, but it could be a generic manifold endowed with a generic metric tensor $g$ and a generic skewness tensor $T$.
This is particularly useful with respect to quantum mechanics, where quantum states are probability amplitudes and not genuine probability distributions, and where, for invertible mixed states, there is an infinite number of possible generalizations of the Fisher-Rao metric\footnote{These are metric tensors satisfying the so-called monotonicity property, i.e., the scalar product they induce on tangent vectors does not increase under the action of completely positive trace-preserving (CPTP) maps.} (\cite{petz-monotone_metrics_on_matrix_spaces}, \cite{petz-quantum_information_theory_and_quantum_statistics}). 

To keep the article as self-contained as possible, we briefly recall the main points of Hamilton-Jacobi theory (for a more detailed formulation of the problem we refer to \cite{marmo_morandi_mukunda-a_geometrical_approach_to_the_hamilton-jacobi_form_of_dynamics_and_its_generalizations, arnold-mathematical_methods_of_classical_mechanics, abraham_marsden-foundations_of_mechanics}).
In the variational formulation of dynamics \cite{lanczos-the_variational_principles_of_mechanics}, the solutions of the equations of motion are expressed as the critical points of the action functional:

\begin{equation}\label{eqn: action}
I\left(\gamma\right)=\int_{t_{\mathrm{in}}}^{t_{\mathrm{fin}}}\,\lag\left(\gamma\,,\dot{\gamma}\right)\,\mathrm{d}t\,,
\end{equation}
where $\gamma$ are curves with fixed extreme points $q(t_{\mathrm{in}})=q_{\mathrm{in}}$ and $q(t_{\mathrm{fin}})=q_{\mathrm{fin}}$, and $\lag$ is the Lagrangian function of the system.
The evaluation of the action functional on a critical point $\gamma_{c}$ gives a two-point function:

\begin{equation}\label{eqn: H-J solution}
S\left(q_{\mathrm{in}}\,,q_{\mathrm{fin}}\right)=I(\gamma_{c})\,,
\end{equation}
which is known in the literature as the Hamilton characteristic function.
It is a solution of the Hamilton-Jacobi equation for the dynamics:

\begin{equation}
H\left(q\,,\frac{\partial S}{\partial q}\right)=E\,,
\end{equation}
where $H$ is the Hamiltonian function (\cite{carinena_ibort_marmo_morandi-geometry_from_dynamics_classical_and_quantum}) associated with the Lagrangian $\lag$ and $E\in\mathbb{R}$ is a constant.
In particular, $S(q_{\mathrm{in}}\, , q_{\mathrm{fin}})$ is called a complete solution when 

\begin{equation}
det \left| \left| \dfrac{\partial^2 S}{\partial q^j_{\mathrm{in}} \partial q^k_{\mathrm{fin}} } \right|\right| \neq 0 \,. 
\end{equation}
Therefore, $S$ is the generating function of a canonical transformation on the phase space of the system.
Specifically, we have the following relations:

\begin{eqnarray}
\label{cantraf0}
p_{j}^{\mathrm{in}}&=&-\frac{\partial S}{\partial q_{\mathrm{in}}^{j}}\,, \\
p_{j}^{\mathrm{fin}}&=&\frac{\partial S}{\partial q_{\mathrm{fin}}^{j}} \,,
\label{cantrasf1}
\end{eqnarray}
where $\left\{ p^{\mathrm{in}}_{j} \right\}$ (resp. $\left\{ p^{\mathrm{fin}}_{j} \right\}$)  are the canonical momenta associated to $q_{\mathrm{in}}^{j}$'s (resp. $q_{\mathrm{fin}}^{j}$'s).

\vsp 

The fact that $S$ is a two-point function allows us to read the problem of finding a canonical contrast function on a statistical manifold as the Hamilton-Jacobi problem associated with suitable Lagrangian and Hamiltonian functions.
Indeed, consider a statistical manifold $(\mathcal{M},\,  g  , \, T)$, let $\alpha\neq0$ be a real number, and let us define the following Lagrangian function

\be
\lag_{\alpha}(q,v)=\dfrac{1}{2} g_{jk}(q)v^jv^k + \dfrac{\alpha}{6} T_{jkl}(q)v^jv^kv^l \, . 
\label{Lagrangian}
\ee      
Our claim is that a complete solution $S_{\alpha}$ of the Hamilton-Jacobi equation associated with this Lagrangian is a potential function for our statistical manifold in the sense that it allows to recover the geometric structure of the manifold as follows:

\be
\left.\frac{\partial^{2}\,S_{\alpha}}{\partial q^{j}_{\mathrm{fin}}\partial q^{k}_{\mathrm{in}}}\right|_{q_{\mathrm{in}}=q_{\mathrm{fin}}}=-g_{jk}\,,
\label{metric2}
\ee

\be
\left. \dfrac{\partial ^3 S_{\alpha}}{\partial q^{l}_{\mathrm{in}} \partial q^{k}_{\mathrm{in}} \partial q^{j}_{\mathrm{fin}} }\right|_{q_{\mathrm{in}}=q_{\mathrm{fin}}} - \left. \dfrac{\partial ^3 S_{\alpha}}{\partial q^{l}_{\mathrm{fin}} \partial q^{k}_{\mathrm{fin}} \partial q^{j}_{\mathrm{in}} }\right|_{q_{\mathrm{in}}=q_{\mathrm{fin}}} = 2\alpha T_{jkl}\,.
\label{tensor2}
\ee

Equation \eqref{tensor2} is slightly different from equation \eqref{tensor}, consequently, we have chosen the name potential function instead of contrast function for $S_{\alpha}$ because, as we shall see, $S_{\alpha}$ allows us to recover the geometrical structures of the statistical manifold. 
Notice further that $S_{\alpha}$ does not need to be positive semidefinite, while a contrast function must be so.

\begin{remark}
In the following, we will use the term ``contrast function'' for a two-point function satisfying the conditions given in equations \eqref{defining condition 1} and \eqref{defining condition 2}, see paragraph after Remark 1.
We will use the term ``potential function'' for a generic two-point function by means of which it is possible to recover the geometrical structure of a statistical manifold $(\mathcal{M},\,  g  , T)$ using a suitably defined procedure, such as that given by equations \eqref{metric2} and \eqref{tensor2}.
\end{remark}

Note that it is possible to write the Lagrangian \eqref{Lagrangian} in intrinsic form as follows:

\be
\lag_{\alpha}= \lag_g + \frac{1}{3} L_{_{\alpha}\Gamma}\lag_g \,,
\label{lag}
\ee
where $\lag_g$ is the metric Lagrangian associated with $g$ and $_{\alpha}\Gamma$ is the second order vector field \cite{marmo_ferrario_lovecchio_morandi_rubano-the_inverse_problem_in_the_calculus_of_variations_and_the_geometry_of_the_tangent_bundle} associated to the affine connection $\nabla^{\alpha}$ \eqref{alpha-connections}. By looking at this expression we can notice that this Lagrangian can be considered as a sort of first-order approximation of a more complete function $\widetilde{\lag_{\alpha}}$ which also includes all successive Lie derivatives with respect to the vector field $\Gamma$, shortly:


\be
\widetilde{\lag_{\alpha}}= \mathrm{e}^{\tau\, L_{_{\alpha}\Gamma}}\lag_g=\left(\Phi_{\tau}^{\alpha}\right)^{*}\lag_{g} \,,
\label{lagmod}
\ee  
where $\Phi_{\tau}^{\alpha}$ is the flow of the second order vector field $_{\alpha}\Gamma$ on the tangent bundle $T\mathcal{M}$ of the statistical manifold $\mathcal{M}$.
However, only \eqref{lag} contributes to the determination of metric and skewness tensors, as we will prove in the following.

\vsp

According to equation \eqref{cantraf0} we have $\frac{\partial S_{\alpha}}{\partial q_{\mathrm{in}}^{j}}=-p_{j}^{\mathrm{in}}$. Furthermore, the momenta $p_{j}$ can be expressed in terms of the Lagrangian function as:

\begin{equation}
p_{j}=\frac{\partial \lag}{\partial v^{j}}\,,
\end{equation}
in particular, for our Lagrangian we get:

\begin{equation}
\label{canmom}
p_{j}=g_{jk}(q)v^{k} + \frac{\alpha}{2}\,T_{jkl}(q)v^{k}v^{l}\,.
\end{equation}
From this, it follows that:

\begin{equation}
\dfrac{\partial S_{\alpha}}{\partial q_{\mathrm{in}}^{j} }=-p_{j}^{\mathrm{in}}=-g_{jk}(q_{\mathrm{in}})v_{\mathrm{in}}^{k} - \frac{\alpha}{2}\,T_{jkl}(q_{\mathrm{in}})v_{\mathrm{in}}^{k}v_{\mathrm{in}}^{l}\,, 
\end{equation}
where the $v^{j}_{in}$'s must be expressed in terms of the initial and final positions.
The link between initial and final positions $(q_{\mathrm{in}},q_{\mathrm{fin}})$ and the initial velocity $(v_{\mathrm{in}})$ is provided by the dynamical trajectories $\gamma_c$ associated with the Lagrangian $\lag$. The Euler-Lagrange equations associated with $\lag$ are:

\be
\left(g_{jk}(q) + \alpha T_{jkl}v^{l}\right)\dot{v}^{k}\,=\,-\,_{g}\Gamma_{jkl}v^{l}v^{k} - \frac{\alpha}{6}\left(\frac{\partial T_{jkl}}{\partial q^{m}} + \frac{\partial T_{jlm}}{\partial q^{k}} + \frac{\partial T_{jkm}}{\partial q^{l}} - \frac{\partial T_{klm}}{\partial q^{j}}\right)v^{k}v^{l}v^{m}\,,
\label{eulLag}
\ee
where $v^{j}=\frac{\mathrm{d} q^{j}}{\mathrm{d}t}$ and $\dot{v}^{j}=\frac{\mathrm{d} v^{j}}{\mathrm{d}t}$.
A series expansion of $\gamma_{c}(t)=(q^{1}(t)\,,\cdots\,,q^{n}(t))$ around $t=0$ yields:

\begin{equation}
q^{j}(t)= q_{\mathrm{in}}^{j} + t\,\left.\frac{\mathrm{d} q^{j}}{\mathrm{d}t}\right|_{t=0} + \frac{t^{2}}{2}\,\left.\frac{\mathrm{d}^{2} q^{j}}{\mathrm{d}t^{2}}\right|_{t=0} +  \mathcal{O}(t^{3})\,.
\end{equation}  
We know that $v_{\mathrm{in}}=\left.\frac{\mathrm{d} q^{j}}{\mathrm{d}t}\right|_{t=0}$, so that naming $q_{\mathrm{fin}}^{j}:=q^{j}(1)$, we can write:

\be
v_{\mathrm{in}}^{j}=q_{\mathrm{fin}}^{j} -  q_{\mathrm{in}}^{j} - \frac{1}{2}\,\left.\frac{\mathrm{d} v_{\mathrm{in}}^{j}}{\mathrm{d}t}\right|_{t=0}\,,
\label{v_in}
\ee
where higher order terms in the expansion have been neglected.
Since $v_{\mathrm{in}}$ is function of $q_{\mathrm{in}},q_{\mathrm{fin}}$, we can express the derivatives with respect to $q_{\mathrm{fin}}$ in terms of the derivatives of $v_{\mathrm{in}}$ and viceversa.
Indeed\footnote{Note that, at this level, formula \eqref{eqn: utile per kullback-leibler 1} and \eqref{eqn: utile per kullback-leibler 2} are valid for every dynamical curve $\gamma$.}:

\begin{eqnarray}\label{eqn: utile per kullback-leibler 1}
\frac{\partial }{\partial q_{\mathrm{fin}}^{j}} & = &\frac{\partial v_{\mathrm{in}}^{k}}{\partial q_{\mathrm{fin}}^{j}}\,\frac{\partial}{\partial v_{\mathrm{in}}^{k}}\, , \\ \label{eqn: utile per kullback-leibler 2}
\frac{\partial^{2} }{\partial q_{\mathrm{fin}}^{k}\partial q_{\mathrm{fin}}^{j}} & = &\frac{\partial^{2} v_{\mathrm{in}}^{l}}{\partial q_{\mathrm{fin}}^{k} \partial q_{\mathrm{fin}}^{j}}\,\frac{\partial}{\partial v_{\mathrm{in}}^{l}} + \frac{\partial v_{\mathrm{in}}^{l}}{\partial q_{\mathrm{fin}}^{j}}\,\frac{\partial v_{\mathrm{in}}^{r}}{\partial q_{\mathrm{fin}}^{k}}\,\frac{\partial}{\partial v_{\mathrm{in}}^{r}}\,\frac{\partial}{\partial v_{\mathrm{in}}^{l}} \, ,
\end{eqnarray}
and we need to evaluate these expressions on the diagonal $q_{\mathrm{fin}}=q_{\mathrm{in}}$.
Note that the condition $q_{\mathrm{fin}}=q_{\mathrm{in}}$ is equivalent to the fact that the dynamical trajectory is $q^{j}(t)=q_{\mathrm{in}}^{j}$, and thus, according to the equations of motion, this corresponds to $v_{\mathrm{in}}=0$.

Equation $\eqref{eulLag}$ can be written as follows:

\begin{equation}
 \dot{v}^{l}\,=\, - \alpha T^{l}_{jk}v^{k}\dot{v}^{j} -\,_{g}\Gamma^{l}_{kj}v^{j}v^{k} - \frac{\alpha}{6}\,g^{lj}\left(\frac{\partial T_{jkr}}{\partial q^{m}} + \frac{\partial T_{jrm}}{\partial q^{k}} + \frac{\partial T_{jkm}}{\partial q^{r}} - \frac{\partial T_{krm}}{\partial q^{j}}\right)v^{k}v^{r}v^{m}\,.
\end{equation}

If we suppose that $\dot{v}^j$ is an analytic function of $\left\lbrace v^j \right\rbrace$ in a neighbourhood of $v^j=0$, we can write 

\begin{equation}
\dot{v}^k = \sum_{m=0}^{\infty} \: \sum_{j_1, \cdots ,j_m =1}^n a^k_{j_1\cdots j_m}v^{j_1}\cdots v^{j_m}\,.
\end{equation}
By inserting this expression into equation $\eqref{eulLag}$ we get the coefficients

\begin{eqnarray}
a^k_0 &=& 0\,, \\
a^k_{j_1} &=& 0\,, \\
a^k_{j_1j_2} &=& -_{g}\Gamma^{k}_{j_1 j_2}\,,
\end{eqnarray}
and so on.

Therefore $\dot{v}^j$ is a function of order $\mathcal{O}(|v|^2)$ and higher order derivatives $\left\lbrace \ddot{v}^j , \, \cdots \right\rbrace$ will be at least of order $\mathcal{O}(|v|^3)$. We can now put these results into equation $\eqref{v_in}$ to obtain

\begin{equation}\label{eqn: utile per kullback-leibler 5}
v_{\mathrm{in}}^j = q_{\mathrm{fin}}^j - q_{\mathrm{in}}^j + \frac{1}{2} {_g\Gamma^j_{kl}v_{\mathrm{in}}^kv_{\mathrm{in}}^l} + \mathcal{O}(|v|^3)\,.
\end{equation} 

Deriving this expression with respect to $q_{\mathrm{fin}}$ and then evaluating it at $v_i=0$ we have 

\begin{align}\label{eqn: utile per kullback-leibler 3}
&\left.\frac{\partial v_i^k }{\partial q_{\mathrm{fin}}^{j}}\right|_{q_{\mathrm{in}}=q_{\mathrm{fin}}}= \: \delta_j^k\,, \\ \label{eqn: utile per kullback-leibler 4}
&\left.\frac{\partial^{2}v^l_i }{\partial q_{\mathrm{fin}}^{k}\partial q_{\mathrm{fin}}^{j}}\right|_{q_{\mathrm{in}}=q_{\mathrm{fin}}} = \: _g\Gamma^l_{jk}\,.
\end{align}
    
Eventually we get:

\begin{align}
\label{Hessian}
&\left. \dfrac{\partial ^2 S_{\alpha}}{\partial q_{\mathrm{fin}}^k \partial q_{\mathrm{in}}^j } \right| _{q_{\mathrm{in}}=q_{\mathrm{fin}}} = -\left.\frac{\partial p_{j}^{\mathrm{in}}}{\partial v_{\mathrm{in}}^{k}}\right|_{v_{\mathrm{in}}=0}  =  -g_{jk} \,,\\
& \left. \dfrac{\partial ^3 S_{\alpha}}{\partial q_{\mathrm{fin}}^l \partial q_{\mathrm{fin}}^k \partial q_{\mathrm{in}}^j }\right|_{q_{\mathrm{in}}=q_{\mathrm{fin}}} = - \,\left._{g}\Gamma^{r}_{kl}\,\frac{\partial p_{j}^{\mathrm{in}}}{\partial v_{\mathrm{in}}^{r}}\right|_{v_{\mathrm{in}}=0}  - \left.\frac{\partial ^2 p_{j}^{\mathrm{in}}}{\partial v_{\mathrm{in}}^{l} \partial v_{\mathrm{in}}^{k}}\,\right|_{v_{\mathrm{in}}=0} =\:  -_{g}\Gamma_{jkl} - \alpha T_{jkl}\,.
\label{skew}
\end{align}

From them it follows immediately that the metric tensor is derived from the metric Lagrangian only, whereas information about the connection depends on the ``interaction term'', as it should be since quadratic terms alone cannot contribute to third order derivatives. In particular, when $\alpha=0$ we get the Christoffel symbols of the Levi-Civita connection associated to the metric $g$.

In order to extract the symmetric tensor from our potential function we need to take derivatives in a different order, according to equation $\eqref{tensor2}$. The main difference with respect to our previous description is the fact that 

\begin{equation}
\dfrac{\partial S_{\alpha}}{\partial q^{j}_{\mathrm{fin}} } = p_{j}^{\mathrm{fin}}\,,
\end{equation}
which is the canonical momentum at the extreme $\gamma(t=1)=q_{\mathrm{fin}}$. 
 
Following the procedure just outlined, we have to express the dependence of $v_{\mathrm{fin}}$ on the variables $(q_{\mathrm{in}}, \, q_{\mathrm{fin}})$. This relation is provided by the dynamics, which is reversible. Then it follows that 

\begin{equation}
v^{j}_{\mathrm{fin}}= q^{j}_{\mathrm{fin}} - q^{j}_{\mathrm{in}} - \frac{1}{2} \,  _{g}\Gamma^{j}_{kl} v^{k}_{\mathrm{fin}}v^{l}_{\mathrm{fin}} + \mathcal{O}(|v|^3)\,.
\end{equation} 

Eventually we get 

\begin{align}
\label{Hes2}
&\left. \dfrac{\partial ^2 S_{\alpha}}{\partial q^{k}_{\mathrm{in}} \partial q^{j}_{\mathrm{fin}} } \right| _{q_{\mathrm{in}}=q_{\mathrm{fin}}} =  -\left.\frac{\partial p_{j}^{\mathrm{fin}}}{\partial v_{\mathrm{fin}}^{k}}\right|_{v_{\mathrm{fin}}=0}  =  -g_{jk}\,, \\
& \left. \dfrac{\partial ^3 S_{\alpha}}{\partial q^{l}_{\mathrm{in}} \partial q^{k}_{\mathrm{in}} \partial q^{j}_{\mathrm{fin}} }\right|_{q_{\mathrm{in}}=q_{\mathrm{fin}}} = -\,\left._{g}\Gamma^{r}_{kl}\,\frac{\partial p_{j}^{\mathrm{fin}}}{\partial v^{r}_{\mathrm{fin}}}\right|_{v_{\mathrm{fin}}=0} + \left.\frac{\partial ^2 p_{j}^{\mathrm{fin}}}{\partial v^{l}_{\mathrm{fin}} \partial v^{k}_{\mathrm{fin}}}\,\right|_{v_{\mathrm{fin}}=0} =\: - _{g}\Gamma_{jkl} + \alpha T_{jkl}\,.
\end{align}
From equations $\eqref{metric2}$ and $\eqref{tensor2}$ it follows that $S_{\alpha}$ is actually a potential function for the statistical manifold $(M,\,  g  , T)$.
Note that $S_{\alpha}$ needs not to be positive as a contrast function would be, however this is not an obstruction in determining metric and skewness tensors. Indeed it is sufficient that $S_{\alpha}$ has a local extreme on the diagonal $q_{\mathrm{in}}=q_{\mathrm{fin}}$ (or $v=0$), and it is true in our case as shown above, by calculating the Hessian matrices $\eqref{Hessian}$, \eqref{Hes2}. Furthermore, as already noticed, one could also think to add other interaction terms to the basic Lagrangian \eqref{lag}, and by a suitable choice of the coupling constants it is possible to make the contrast function positive definite. 

Furthermore, we want to stress that, from the purely mathematical point of view, the informations contained in the metric tensor $g$ and the skewness tensor $T$ are completely uncorrelated with the fact that $\mathcal{M}$ is actually a manifold of probability distributions as it is the case in classical information geometry.
The Hamilton-Jacobi approach to potential functions introduced here makes reference only to the geometrical properties of $\mathcal{M}$ encoded in $g$ and $T$, and no reference to the fact that $\mathcal{M}$ is a manifold of probability distributions is needed.
Therefore, this approach works perfectly well when $\mathcal{M}$ is not a manifold of probability distributions, as it is the case for quantum information geometry, where the focus is on quantum states, i.e., probability amplitudes, and the relevant statistical manifolds are manifolds of quantum states.

\section{Features of the Hamilton-Jacobi approach}\label{sec: discussion}

Here we will discuss some features of the Hamilton-Jacobi approach to potential functions described in the previous section.

Let us start pointing out the connection between the potential function $S_{\alpha}$ defined here and the canonical contrast function defined on self-dual statistical manifolds (\cite{amari_nagaoka-methods_of_information_geometry}).
We recall that these are statistical manifolds for which the symmetric tensor $T$ identically vanishes, so that the only connection at our disposal is the self-dual Levi-Civita connection $\nabla_{g}$ associated with the metric $g$.

For self-dual manifolds, a canonical contrast function $S_{d}$ exists which is given by:

\be\label{eqn: contrast function for self-dual manifolds}
S_{d}(\mathbf{q}_{\mathrm{in}}\,,\mathbf{q}_{\mathrm{fin}})=\frac{1}{2}\,d^{2}(\mathbf{q}_{\mathrm{in}}\,,\mathbf{q}_{\mathrm{fin}})\,,
\ee
where $d^{2}(q_{\mathrm{in}}\,,q_{\mathrm{fin}})$ is the square of the Riemannian geodesic distance associated with the metric $g$ on $\mathcal{M}$.

Applying our procedure to the case of self-dual manifolds, it is clear that the potential function we obtain is precisely the canonical contrast function $S_{d}$ defined above.
To see this, recall that the metric Lagrangian $\lag_{g}$ associated with the metric tensor $g$, and all of its functions $F(\lag_{g})$ with $F$ analytic, give rise to the same dynamical trajectories (\cite{marmo_ferrario_lovecchio_morandi_rubano-the_inverse_problem_in_the_calculus_of_variations_and_the_geometry_of_the_tangent_bundle}).
Furthermore, $\lag_{g}$ and $F(\lag_{g})$ are all constants of the motion for the dynamics.
Denoting with $\gamma_{c}$  the geodesic connecting $\mathbf{q}_{\mathrm{in}}=\gamma_{c}(0)$ and $\mathbf{q}_{\mathrm{fin}}=\gamma_{c}(1)$, this implies that:

\be
I_{F(\lag_{g})}(\mathbf{q}_{\mathrm{in}}\,,\mathbf{q}_{\mathrm{fin}}):=\int_{t_{\mathrm{in}}}^{t_{\mathrm{fin}}}\,F(\lag_{g})\,\mathrm{d}t=\left.F(\lag_{g})\right|_{\gamma_{c}(t)}\,.
\ee
Now, the Riemannian geodesic distance $d^{2}(q_{\mathrm{in}}\,,q_{\mathrm{fin}})$ is given by:

\be
d(\mathbf{q}_{\mathrm{in}}\,,\mathbf{q}_{\mathrm{fin}})=\int_{t_{\mathrm{in}}}^{t_{\mathrm{fin}}}\,\sqrt{2\lag_{g}}\,\mathrm{d}t\,,
\ee
and thus:

$$
d^{2}(\mathbf{q}_{\mathrm{in}}\,,\mathbf{q}_{\mathrm{fin}})=\left(\int_{t_{\mathrm{in}}}^{t_{\mathrm{fin}}}\,\sqrt{2\lag_{g}}\,\mathrm{d}t\right)^{2}=\left.\left(\sqrt{2\lag_{g}}\right)^{2}\right|_{\gamma_{c}(t)} =
$$
\be 
\;\;\;\;\;\;\;\;\;\;\;\;\;\;\;\;\;\;\;\;\;\;\;\;\;=\left.2\lag_{g}\right|_{\gamma_{c}(t)}=2\int_{t_{\mathrm{in}}}^{t_{\mathrm{fin}}}\,\lag_{g}\,\mathrm{d}t=2\,I_{\lag_{g}}(\mathbf{q}_{\mathrm{in}}\,,\mathbf{q}_{\mathrm{fin}})
\ee
Equations  \eqref{eqn: action}, \eqref{eqn: H-J solution}, and \eqref{Lagrangian}, with $T=0$, allow us to immediately realize that the Hamilton characteristic function associated with the Lagrangian $\lag_{\alpha}=\lag_{g}$ is precisely the canonical contrast function of equation \eqref{eqn: contrast function for self-dual manifolds}.

\vsp

A physically interesting example of self-dual statistical manifold is given by the manifold $\mathcal{P}(\mathcal{H})$ of pure states of quantum mechanics.
As it is shown in \cite{wootters-statistical_distance_and_hilbert_space}, a meaningful notion of statistical distance between pure states can be defined by means of the concepts of distinguishability  and statistical fluctuations in the outcomes of measurements.
It turns out that this two-point function on pure states coincides with the Riemannian geodesic distance associated with the Fubini-Study metric, so that the statistical structure  determined by this two-point function makes $\mathcal{P}(\mathcal{H})$ a self-dual manifold.
Consequently, we can apply our procedure and conclude that the statistical distance  introduced by Wotters coincides with the Hamilton principal function associated with the metric Lagrangian of the Fubini-Study metric on $\mathcal{P}(\mathcal{H})$.
The relevance of the statistical structure on $\mathcal{P}(\mathcal{H})$ emerging from Wotter's statistical distance is enforced by the results of \cite{ay_tuschmann-dually_flat_manifolds_and_global_information_geometry, ay_tuschmann-duality_versus_flatness_in_quantum_information_geometry}, where it is shown that the set of pure states of quantum mechanics does not admit a dually flat statistical structure.

\vsp

Now, consider the statistical manifold $(\mathcal{M}\,,g\,,T)$, where $\mathcal{M}=\mathbb{R}^{+}$, $g=\frac{1}{\xi^{2}}$, $_{g}\Gamma=-\frac{1}{\xi}$ is the Christoffel symbol of the Levi-Civita connection and $T =-\frac{2}{\xi^3} $.
This manifold arises as the statistical manifold associated to the exponential distributions

\be
\mathit{p}(x\,,\xi)=\xi\,\mathrm{e}^{-x \xi}\qquad \xi,x>0\,.
\ee
The metric $g$ and the tensor $T$ are then obtained by:

\be
g=\int_{0}^{+\infty}\,\mathit{p}(x\,,\xi)\,\left(\frac{\mathrm{d}\log(\mathit{p})}{\mathrm{d}\xi}\right)^{2}\,\mathrm{d}x\,,
\ee

\be
T=\int_{0}^{+\infty}\,\mathit{p}(x\,,\xi)\,\left(\frac{\mathrm{d}\log(\mathit{p})}{\mathrm{d}\xi}\right)^{3}\,\mathrm{d}x\,.
\ee

The ``deformed'' Lagrangian function $\lag$ with respect to the connection $\nabla$ reads:

\be\label{eqn: lagrangian 1-d exponential distributions}
\lag_{\alpha}=\frac{v^{2}}{2\xi^{2}} - \frac{\alpha}{3}\frac{v^{3}}{\xi^{3}}\,, 
\ee
where $\lag_{g}=\frac{v^{2}}{2\xi^{2}}$ is the metric Lagrangian.
It is clear that $\lag_{\alpha}$ is a function of $\lag_{g}$, specifically, it is $\lag_{\alpha}= \lag_{g} + \frac{2\sqrt{2}\alpha}{3}\left(\lag_{g}\right)^{\frac{3}{2}}$.
Consequently, the solutions of the Euler-Lagrange equations associated with the metric Lagrangian $\lag_{g}$, i.e., the geodesics of $g$, are solutions of the Euler-Lagrange equations associated with the Lagrangian $\lag_{\alpha}$, and the explicit expression of the dynamical trajectories $\gamma_{c}(t)=\xi(t)$ is:

\be
\xi(t)=\xi_{\mathrm{in}}\,\mathrm{e}^{\frac{v_{\mathrm{in}}}{\xi_{\mathrm{in}}}\,t}\,.
\ee
A complete solution of the Hamilton-Jacobi problem for $\lag$ is given by:

\be
I_{\alpha}(\gamma_{c})=\int_{t_{\mathrm{in}}}^{t_{\mathrm{fin}}}\lag_{\alpha}\left(\gamma_{c}(t)\,,\dot{\gamma_{c}}(t)\right)\,\mathrm{d}t\,,
\ee
where the curve $\gamma_{c}$ has fixed extreme points $\xi_{in}=\gamma_{c}(t_{\mathrm{in}})$ and $\xi_{fin}=\gamma_{c}(t_{\mathrm{fin}})$, and integration is performed between $t_{\mathrm{in}}=0$ and $t_{\mathrm{fin}}=1$.
In our case, since the Lagrangian $\lag_{\alpha}$ is a constant of the motion, we have:

\be\label{eqn: Hamilton principal function 1-d exponential distributions}
I_{\alpha}(\gamma_{c})=\frac{v^{2}_{\mathrm{in}}}{2\xi^{2}_{\mathrm{in}}} - \frac{\alpha}{3}\frac{v^{3}_{\mathrm{in}}}{\xi^{3}_{\mathrm{in}}}\,.
\ee
The link between $\xi_{\mathrm{fin}},\xi_{\mathrm{fin}}$ and $v_{\mathrm{in}}$ can easily be extracted form the explicit expression of $\gamma_{c}(t)$, indeed:

\be
v_{\mathrm{in}}=\xi_{\mathrm{in}}\ln\left(\frac{\xi_{\mathrm{fin}}}{\xi_{\mathrm{in}}}\right)\,,
\ee
and thus, the contrast function $S$ reads:

\be\label{eqn: potential function 1-d exponential distributions}
S_{\alpha}(\xi_{\mathrm{in}}\,,\xi_{\mathrm{fin}})=\frac{\ln^{2}\left(\frac{\xi_{\mathrm{fin}}}{\xi_{\mathrm{in}}}\right)}{2} -\frac{\alpha}{3}\ln^{3}\left(\frac{\xi_{\mathrm{fin}}}{\xi_{\mathrm{in}}}\right)\,.
\ee
An explicit calculation gives:

\be
\left. \dfrac{\partial ^2 S_{\alpha}}{\partial \xi_{\mathrm{fin}} \partial \xi_{\mathrm{in}} } \right| _{\xi_{\mathrm{in}}=\xi_{\mathrm{fin}}\equiv \xi} = -\frac{1}{\xi^{2}}\,,
\ee

\be
\left. \dfrac{\partial ^3 S_{\alpha}}{\partial \xi_{\mathrm{fin}} \partial \xi_{\mathrm{fin}} \partial \xi_{\mathrm{in}} }\right|_{\xi_{\mathrm{fin}}=\xi_{\mathrm{in}}\equiv \xi} = \frac{2\alpha + 1}{\xi^{3}}=-_{g}\Gamma - \alpha T\,,
\ee

\be
\left. \dfrac{\partial ^3 S_{\alpha}}{\partial \xi_{\mathrm{fin}} \partial \xi_{\mathrm{in}} \partial \xi_{\mathrm{in}} }\right|_{\xi_{\mathrm{fin}}=\xi_{\mathrm{in}}\equiv \xi} = \frac{1-2\alpha}{\xi^{3}}=-_{g}\Gamma + \alpha T\,,
\ee
showing that $S_{\alpha}$ is a potential function for the statistical manifold $(\mathcal{M}\,,g\,,T)$ of exponential distributions.

\vsp

The statistical structure of $\mathcal{M}$ can be alternatively derived  starting with the Kullback-Leibler divergence function $S_{KL}$:

\be\label{eqn: K-L divergence 1d example}
S_{KL}(\xi_{\mathrm{in}}\,,\xi_{\mathrm{fin}})=\int_{0}^{+\infty}\,\mathit{p}(x\,,\xi_{\mathrm{in}})\ln\left(\frac{\mathit{p}(x\,,\xi_{\mathrm{in}})}{\mathit{p}(x\,,\xi_{\mathrm{fin}})}\right)\mathrm{d}x=\ln\left(\frac{\xi_{\mathrm{in}}}{\xi_{\mathrm{fin}}}\right) + \frac{\xi_{\mathrm{fin}}}{\xi_{\mathrm{in}}} - 1\,.
\ee
As it is clear, the potential function $S_{\alpha}$ in equation \eqref{eqn: potential function 1-d exponential distributions} does not coincide with the Kullback-Leibler divergence.
This is not surprising since, for a given statistical manifold, there are infinite many potential (contrast) functions generating the same statistical structure.
However, we will now show that it is possible to read the Kullback-Leibler divergence $S_{KL}$ as the Hamilton principal function $I_{KL}$ associated with a suitably defined Lagrangian.
At this purpose, let us perform the following diffeomorphism between $\mathcal{M}=\mathbb{R}_{+}$ and $\mathcal{N}=\mathbb{R}$:

\be\label{eqn: diffeomorphism 1d exponential distribution}
\xi\mapsto\, y=\ln(\xi)\,.
\ee
This diffeomorphism gives rise to a diffeomorphism between  $\mathcal{M}\times\mathcal{M}$ and $\mathcal{N}\times\mathcal{N}$:

\be
\left(\xi_{\mathrm{in}}\,,\xi_{\mathrm{fin}}\right)\mapsto\,\left(y_{\mathrm{in}}=\ln(\xi_{\mathrm{in}})\,,y_{\mathrm{fin}}=\ln(\xi_{\mathrm{fin}})\right)\,.
\ee
The Kullback-Leibler divergence $S_{KL}$ becomes:

\be\label{eqn: kullback-leibler 1d on R}
S_{KL}\left(y_{\mathrm{in}}\,,y_{\mathrm{fin}}\right)= \mathrm{e}^{(y_{\mathrm{fin}} - y_{\mathrm{fin}})} -\left(y_{\mathrm{fin}}-y_{\mathrm{in}}\right) - 1\,.
\ee
Now, consider the following Lagrangian on $T\mathcal{N}$:

\be
\lag_{KL}(y\,,u)=\mathrm{e}^{u} - u - 1\,,
\ee
and let us calculate the Hamilton principal function associated with $\lag_{KL}$.
Since $\lag_{KL}$ depends only on the velocity coordinate $u$, it is an alternative Lagrangian for the $1$-dimensional free-particle on $\mathcal{N}=\mathbb{R}$ (\cite{marmo_ferrario_lovecchio_morandi_rubano-the_inverse_problem_in_the_calculus_of_variations_and_the_geometry_of_the_tangent_bundle}).
Consequently, the dynamical trajectories of the system coincide with the geodesics of the Euclidean metric, that is, they are straight lines $\gamma_{c}(t)=v_{\mathrm{in}}t + y_{\mathrm{in}}$.
Setting $t_{\mathrm{in}}=0\,,t_{\mathrm{fin}}=1$, the connection between $y_{\mathrm{in}}\,,y_{\mathrm{fin}}$ and $v_{\mathrm{in}}$ is easily seen to be $v_{\mathrm{in}}=y_{\mathrm{fin}}-y_{\mathrm{in}}$.
Furthermore, $\lag_{KL}$ is a constant of the motion, and thus, it can be brought out from the integral defining the Hamilton principal function:

\be\label{eqn: kullback-leibler 1d on tangent bundle}
I_{KL}\left(y_{\mathrm{in}}\,,y_{\mathrm{fin}}\right)=\lag_{KL}\left(y_{\mathrm{in}}\,,v_{\mathrm{in}}(y_{\mathrm{in}}\,,y_{\mathrm{fin}})\right)= \mathrm{e}^{(y_{\mathrm{fin}} - y_{\mathrm{fin}})} - \left(y_{\mathrm{fin}}-y_{\mathrm{in}}\right) -1\,.
\ee
Confronting equations \eqref{eqn: kullback-leibler 1d on R} and \eqref{eqn: kullback-leibler 1d on tangent bundle}, we conclude that the Hamilton principal function associated with $\lag_{KL}$ is precisely the Kullback-Leibler divergence $S_{KL}$ as claimed.

Now we will point out an interesting connection between the Lagrangian $\lag_{KL}$ associated with the Kullback-Leibler divergence of equation \eqref{eqn: K-L divergence 1d example} and the Lagrangian $\lag_{\alpha}$ associated with the potential function of equation \eqref{eqn: potential function 1-d exponential distributions}.
At this purpose, let us apply the tangent lift of the diffeomorphism given by equation \eqref{eqn: diffeomorphism 1d exponential distribution} to the Lagrangian $\lag_{\alpha}$ of equation \eqref{eqn: lagrangian 1-d exponential distributions}:

\be\label{eqn: lagrangian 1-d exponential distribution on R}
\lag_{\alpha}(y\,,u)=\frac{u^{2}}{2} - \frac{\alpha}{3}\,u^{3}\,.
\ee
Now, let us perform a series expansion of $\lag_{KL}$ around $u=0$:

\be\label{eqn: series expansion of K-L lagrangian 1-d exponential distribution on R}
\lag_{KL}(y\,,u)=\frac{u^{2}}{2} + \frac{u^{3}}{6} +  \mathcal{O}(u^{4})\,. 
\ee

Confronting equations \eqref{eqn: lagrangian 1-d exponential distribution on R} and \eqref{eqn: series expansion of K-L lagrangian 1-d exponential distribution on R} we immediately see that, upon taking $\alpha=-\frac{1}{2}$, the Lagrangian $\lag_{\alpha}$ is precisely the third order approximation of $\lag_{KL}$.

We can push this line of reasoning a little further, and show that, if the Kullback-Leibler divergence $S_{KL}$ generating the statistical structure of a statistical manifold $(\mathcal{M}\,,g\,,T)$ is the Hamilton principal function associated with a Lagrangian $\lag_{KL}$, then the Lagrangian $\lag_{\alpha}$ we have proposed here is the third order approximation of $\lag_{KL}$ up to a constant factor, and  provided we choose $\alpha=-\frac{1}{2}$.

Let us consider a statistical manifold $(\mathcal{M}\,,g\,,T)$ and the Kullback-Leibler divergence $S_{KL}$ generating the statistical structure of $\mathcal{M}$.
Let us assume that $S_{KL}$ admits a Lagrangian $\lag_{KL}$ such that $S_{KL}$ is the Hamilton principal function associated with $\lag_{KL}$.
Assuming $\lag_{KL}$ analytic in $\mathbf{v}$ as in the previous section, an expansion of $\lag_{KL}$ in a power series of the velocity vector $\mathbf{v}$ around $\mathbf{v}=0$ gives:

$$
\lag_{KL}=\left. \lag_{KL}\right|_{\mathbf{v}=0} + \left. \frac{\partial \lag_{KL}}{\partial v^{j}}\right|_{\mathbf{v}=0}\,v^{j} + \left. \frac{\partial^{2} \lag_{KL}}{\partial v^{j}\partial v^{k}}\right|_{\mathbf{v}=0}\,\frac{v^{j}v^{k}}{2!} + \left. \frac{\partial^{3} \lag_{KL}}{\partial v^{j}\partial v^{k}\partial v^{l}}\right|_{\mathbf{v}=0}\,\frac{v^{j}v^{k}v^{l}}{3!} + \mathcal{O}(\mathbf{v}^{4})\,.
$$
We will now examine the terms of the expansion up to the third order.

Concerning the first order term, we recall that $S_{KL}$ must have a minimum on the diagonal $\mathbf{q}_{\mathrm{in}}=\mathbf{q}_{\mathrm{fin}}$ of $\mathcal{M}\times\mathcal{M}$.
Therefore, recalling equations \eqref{cantraf0} and \eqref{eqn: sufficient condition for second derivative to form a tensor}, and expressing the derivatives with respect to $v^{j}$ by means of the derivatives with respect to $q_{\mathrm{fin}}^{j}$,  we have:

\be\label{eqn: first order term of expansion of k-l lagrangian}
\left. \frac{\partial \lag_{KL}}{\partial v^{j}}\right|_{\mathbf{v}=0}=-\left. \frac{\partial S_{KL}}{\partial q^{j}_{\mathrm{in}}}\right|_{\mathbf{q}_{\mathrm{in}}=\mathbf{q}_{\mathrm{fin}}}=0\,.
\ee
This equation implies that $\lag_{KL}$ is at least of second order in $\mathbf{v}$.
Consequently, the analysis of the previous section for the functional dependence between $\mathbf{v}$  and $\mathbf{q}_{\mathrm{fin}}$ stemming from the Euler-Lagrange equations can be analogously repeated in order to give the first order relation:

\be
\left.\frac{\partial q^{j}_{\mathrm{fin}}}{\partial v^{j}}\right|_{\mathbf{v}=0}=\delta^{j}_{k}\,.
\ee
This result, together with equations \eqref{eqn: sufficient condition for second derivative to form a tensor}, \eqref{metric} and \eqref{eqn: first order term of expansion of k-l lagrangian}, allows us to see that the second order term of $\lag_{KL}$ becomes:

\be
\left. \frac{\partial^{2} \lag_{KL}}{\partial v^{j}\partial v^{k}}\right|_{\mathbf{v}=0}=-\left. \frac{\partial^{2} S_{KL}}{\partial q^{r}_{\mathrm{fin}}\partial q^{j}_{\mathrm{in}}}\right|_{\mathbf{q}_{\mathrm{in}}=\mathbf{q}_{\mathrm{fin}}}\,\left.\frac{\partial q^{r}_{\mathrm{fin}}}{\partial v^{k}}\right|_{\mathbf{v}=0}=g_{jk}
\ee
Now that we have the second-order term of the Lagrangian, we can proceed in the analysis of the functional dependence between $\mathbf{v}$ and $\mathbf{q}_{\mathrm{fin}}$ and find that:

\be
\left.\frac{\partial^{2} q_{\mathrm{fin}}^{l} }{\partial v^{k}\partial v^{j}}\right|_{\mathbf{v}=0} = \: -  _{g}\Gamma^{l}_{jk}\,.
\ee

Consequently, the third order term of the Lagrangian is:

$$
\left. \frac{\partial^{3} \lag_{KL}}{\partial v^{j}\partial v^{k}\partial v^{l}}\right|_{\mathbf{v}=0}=-\left. \frac{\partial^{3} S_{KL}}{\partial q^{r}_{\mathrm{fin}}\partial q^{n}_{\mathrm{fin}}\partial q^{j}_{\mathrm{in}}}\right|_{\mathbf{q}_{\mathrm{in}}=\mathbf{q}_{\mathrm{fin}}}\,\left.\left(\frac{\partial q^{r}_{\mathrm{fin}}}{\partial v^{k}}\,\frac{\partial q^{n}_{\mathrm{fin}}}{\partial v^{l}}\right)\right|_{\mathbf{v}=0} - \left. \frac{\partial^{2} S_{KL}}{\partial q^{r}_{\mathrm{fin}}\partial q^{j}_{\mathrm{in}}}\right|_{\mathbf{q}_{\mathrm{in}}=\mathbf{q}_{\mathrm{fin}}}\,\left.\frac{\partial^{2} q^{r}_{\mathrm{fin}}}{\partial v^{l}\partial v^{k}}\right|_{\mathbf{v}=0}=
$$
\be
=-\left. \frac{\partial^{3} S_{KL}}{\partial q^{j}_{\mathrm{in}}\partial q^{k}_{\mathrm{fin}}\partial q^{l}_{\mathrm{fin}}}\right|_{\mathbf{q}_{\mathrm{in}}=\mathbf{q}_{\mathrm{fin}}} - _{g}\Gamma_{jkl} = -\frac{T_{jkl}}{2}\,,
\ee
where, in the last equality, we have used the fact that:

\be
\left. \frac{\partial^{3} S_{KL}}{\partial q^{j}_{\mathrm{in}}\partial q^{k}_{\mathrm{fin}}\partial q^{l}_{\mathrm{fin}}}\right|_{\mathbf{q}_{\mathrm{in}}=\mathbf{q}_{\mathrm{fin}}}=\; _{g}\Gamma_{jkl} + \frac{T_{jkl}}{2}\,,
\ee
which is formula $4$ of Lemma $2.1$ in \cite{matumoto-any_statistical_manifold_has_a_contrast_function}.

Collecting the results, we can write:

\be
\lag_{KL}=\left. \lag_{KL}\right|_{\mathbf{v}=0} + g_{jk}\,\frac{v^{j}v^{k}}{2!} -\frac{T_{jkl}}{2}\,\frac{v^{j}v^{k}v^{l}}{3!} + \mathcal{O}(\mathbf{v}^{4})\,,
\ee
from which it follows that, choosing $\alpha=-\frac{1}{2}$, the Lagrangian $\lag_{\alpha}$ differs from the third order approximation of $\lag_{KL}$ only by the constant factor $\left. \lag_{KL}\right|_{\mathbf{v}=0}$ as claimed.
It would be very interesting to understand the conditions under which the Kullback-Leibler divergence of a given statistical manifold is the Hamilton principal function of some suitably-defined Lagrangian.

This appears to be relevant in the context of quantum information geometry of mixed states (\cite{amari_nagaoka-methods_of_information_geometry} Chapter $7$).
Here, the quantum counterpart of the Fisher-Rao metric has been studied extensively by Petz and coworkers (\cite{petz-monotone_metrics_on_matrix_spaces}, \cite{petz-quantum_information_theory_and_quantum_statistics} and references therein), who found that there is  an infinite-number of metrics providing a meaningful generalization of the classical Fisher-Rao metric.
Furthermore, unlike the classical case, there is no preferred definition for a skewness tensor $T$.
This means that there is a good amount of freedom in the choice of a statistical structure on the space of quantum states.
The usual way in which a statistical structure is defined is to start considering a generalization of some classical divergence function, and then derive a metric $g$ and a tensor $T$ on the space of quantum states.
Interestingly,  it is possible to use well known examples of quantum relative entropies as quantum divergence functions (\cite{manko_marmo_ventriglia_vitale-metric_on_the_space_of-quantum_states_from_relative_entropy_tomographic_reconstruction}).
As it is clear, in this quantum setting the statistical structure of the space of quantum states depends on the explicit form of the quantum divergence one starts with.
Now, we have seen that the  tensors $g$ and $T$ of a statistical manifold, being it classical or quantum, are completely encoded in the first four terms of the expansion of the Lagrangian associated with a divergence function, therefore, it is reasonable to argue that there must be some other geometrical informations hidden in the divergence function that are not fully captured by $g$ and $T$ alone.
We believe that the dynamical characterization of the quantum divergences stemming from the Hamilton-Jacobi approach outlined here can be frutifully exploited to better understand the relations between the quantum divergences, the statistical structure they induce, and the geometrical structure of the space of quantum states.
However, a careful analysis of these problems requires the use of a more advanced formulation of the Hamilton-Jacobi theory (\cite{marmo_morandi_mukunda-a_geometrical_approach_to_the_hamilton-jacobi_form_of_dynamics_and_its_generalizations, carinena_gracia_martinez_munoz-lecanda_roman-roy-geometric_hamilton-jacobi_theory}), and will be left for further work.

\vsp

Finally, let us comment on the other approaches to the definition of a potential (contrast) function for a statistical manifold $(\mathcal{M}\,,g\,,T)$.
In \cite{ay_amari-a_novel_approach_to_canonical_divergences_within_information_geometry} a canonical contrast function is constructed using the arclength functional $l$ associated with the metric $g$, and the so-called inverse exponential map $Exp^{-1}_{\nabla}$ associated with an affine connection $\nabla$ defined in terms of $T$ and $_{g}\Gamma$.
In this case, information about the geometrical structures of $(\mathcal{M}\,,g\,,T)$ is taken into account separately.
Specifically, the arclength functional $l$ carries information about the metric tensor $g$, while the inverse exponential map $Exp^{-1}_{\nabla}$ carries information on the affine connection $\nabla$, and thus on the symmetric covariant tensor $T$.
The exponential map $Exp_{\nabla}$ provides a correspondence between tangent vectors at a point $m_{\mathrm{in}}\in \mathcal{M}$, and points in $\mathcal{M}$.
Essentially, given a tangent vector $v_{\mathrm{in}}$, the image of the exponential map $Exp(v_{\mathrm{in}})$ is the point $m_{\mathrm{fin}}\in \mathcal{M}$ that is reached from $m_{\mathrm{in}}$ moving along the $\nabla$-geodesic $\gamma_{m_{\mathrm{in}},v_{\mathrm{in}}}$ with initial velocity $v_{\mathrm{in}}$  when $t=1$, that is, $m_{\mathrm{fin}}=Exp_{\nabla}(v_{\mathrm{in}})=\gamma_{m_{\mathrm{in}},v_{\mathrm{in}}}(1)$.
The inverse $Exp^{-1}_{\nabla}$ of this map gives us a correspondence between a point $m_{\mathrm{fin}}\in \mathcal{M}$ and a tangent vector $v_{\mathrm{in}}$ at $m_{\mathrm{in}}\in \mathcal{M}$.
Writing $X(m_{\mathrm{in}}\,,m_{\mathrm{fin}})=Exp^{-1}_{\nabla}(m_{\mathrm{fin}})$, the canonical contrast function $S$ constructed in \cite{ay_amari-a_novel_approach_to_canonical_divergences_within_information_geometry} reads:

\be
S(m_{\mathrm{in}}\,,m_{\mathrm{fin}}):=\int_{0}^{1}\,g\left(X(\gamma(t)\,,m_{\mathrm{fin}})\,,\dot{\gamma}(t)\right)\,\mathrm{d}t\,.
\ee

Following a similar line of reasoning, it is possible to construct another kind of potential function for $(\mathcal{M}\,,g\,,T)$ using the metric Lagrangian $\lag_{g}$ associated with $g$ and the inverse of the exponential map associated with the affine connection $\nabla\equiv\nabla_{\alpha=1}$.
We recall that the affine connection $\nabla\equiv\nabla_{\alpha=1}$ is particularly relevant when dealing with families of exponential probability distributions, since for them it turns  to be flat (\cite{amari_nagaoka-methods_of_information_geometry}).
The equations of motion for the $\nabla$-geodesics are:

\be
\ddot{q}^{j}(t)=-\Gamma^{j}_{kl}(q(t))\dot{q}^{k}(t)\dot{q}^{l}(t)=-\Gamma^{j}_{kl}(q(t))v^{k}(t)v^{l}(t)\,,
\ee
hence, a series expansion of $q^{j}(t)$ around $t=0$ gives:

\be
q^{j}(t)=q^{j}_{\mathrm{in}} + v^{j}_{\mathrm{in}}t -\frac{t^{2}}{2}\Gamma^{j}_{kl}(q_{\mathrm{in}})v^{k}_{\mathrm{in}}v^{l}_{\mathrm{in}} + O(||tv_{\mathrm{in}}||^{3})\,,
\ee
where $q^{j}(0)=q_{\mathrm{in}}$ and $\dot{q}^{j}(0)=v_{\mathrm{in}}$, and the higher order terms are always a product of some functions of the $q^{j}(t)$ with the functions $v^{j}(t)=\dot{q}^{j}(t)$.
Consequently, the exponential map $Exp_{\nabla}(v_{\mathrm{in}})$ reads:

\be
q^{j}_{\mathrm{fin}}=q^{j}(1)=q^{j}_{\mathrm{in}} + v^{j}_{\mathrm{in}} -\frac{1}{2}\Gamma^{j}_{kl}(q_{\mathrm{in}})v^{k}_{\mathrm{in}}v^{l}_{\mathrm{in}} +  O(||v_{\mathrm{in}}||^{3})\,,
\ee
from which we immediately obtain:

\begin{align}
&\left.\frac{\partial q^{j}_{\mathrm{fin}}}{\partial q^{k}_{\mathrm{in}}}\right|_{v_{\mathrm{in}}=0}=\left.\frac{\partial q^{k}_{\mathrm{fin}}}{\partial v^{k}_{\mathrm{in}}}\right|_{v_{\mathrm{in}}=0}=\delta^{j}_{k}\,,\\
&\left.\frac{\partial^{2}q^{j}_{\mathrm{fin}}}{\partial q_{\mathrm{in}}^{k}\partial q_{\mathrm{in}}^{l}}\right|_{v_{\mathrm{in}}=0}=\left.\frac{\partial^{2}q^{j}_{\mathrm{fin}}}{\partial v_{\mathrm{in}}^{k}\partial q_{\mathrm{in}}^{l}}\right|_{v_{\mathrm{in}}=0}=0 \;\;\;\; \left.\frac{\partial^{2}q^{j}_{\mathrm{fin}}}{\partial v_{\mathrm{in}}^{k}\partial v_{\mathrm{in}}^{l}}\right|_{v_{\mathrm{in}}=0}=-\Gamma^{j}_{kl}(q_{\mathrm{in}})\,,
\end{align}
and:

\begin{align}
\label{eqn0: derivative identities for the inverse of the exponential map associated with a spray}
&\left.\frac{\partial v^{j}_{\mathrm{in}}}{\partial q^{k}_{\mathrm{fin}}}\right|_{d}=-\left.\frac{\partial v^{j}_{\mathrm{in}}}{\partial q^{k}_{\mathrm{in}}}\right|_{d}=\delta^{j}_{k}\,, \\
\label{eqn1: derivative identities for the inverse of the exponential map associated with a spray}
&\left.\frac{\partial^{2}v_{\mathrm{in}}^{j}}{\partial q_{\mathrm{in}}^{k}\partial q_{\mathrm{in}}^{l}}\right|_{d}=-\left.\frac{\partial^{2}v_{\mathrm{in}}^{j}}{\partial q_{\mathrm{in}}^{k}\partial q_{\mathrm{fin}}^{l}}\right|_{d}=\Gamma^{j}_{kl}(q_{\mathrm{in}})\,,
\end{align}
where $\left|_{d}\right.$ denotes the evaluation on the diagonal $q_{\mathrm{in}}=q_{\mathrm{fin}}=q$ of $\mathcal{M}\times \mathcal{M}$.
At this point, we define:

\be\label{eqn: coordinates expression of the contrast function with metric lagrangian}
S(q_{\mathrm{in}}\,,q_{\mathrm{fin}}):=\frac{1}{2}g_{jk}(q_{\mathrm{in}})\,v_{\mathrm{in}}^{j}(q_{\mathrm{in}}\,,q_{\mathrm{fin}})v_{\mathrm{in}}^{k}(q_{\mathrm{in}}\,,q_{\mathrm{fin}})\,,
\ee
where the $v_{\mathrm{in}}^{j}(q_{\mathrm{in}}\,,q_{\mathrm{fin}})$ are determined by the inverse of the exponential map.
Then, a careful application of the chain rule and of the relations $\eqref{eqn0: derivative identities for the inverse of the exponential map associated with a spray}$ and $\eqref{eqn1: derivative identities for the inverse of the exponential map associated with a spray}$ to equation $\eqref{eqn: coordinates expression of the contrast function with metric lagrangian}$ gives:

\be
\left.\frac{\partial^{2}S}{\partial q_{\mathrm{fin}}^{k}\partial q_{\mathrm{fin}}^{j}}\right|_{d}=\left.\frac{\partial^{2}S}{\partial q_{\mathrm{in}}^{k}\partial q_{\mathrm{in}}^{j}}\right|_{d}=-\left.\frac{\partial^{2}S}{\partial q_{\mathrm{in}}^{k}\partial q_{\mathrm{fin}}^{j}}\right|_{d}=g_{jk}(q)\,,
\ee
and:

\be
\left.\frac{\partial^{3}S}{\partial q_{\mathrm{fin}}^{l}\partial q_{\mathrm{fin}}^{k} \partial q_{\mathrm{in}}^{j}}\right|_{d}= - _{g}\Gamma_{jkl} + T_{jkl}\,,
\ee
from which it follows that $S$ is a potential function for $(\mathcal{M}\,,g\,,T)$ as claimed.

\vsp

As it is clear, the non-uniqueness of the potential (contrast) function $S$ implies that the ``inverse problem'' has many alternative solutions and all solutions are to be considered permissible.
However, the Hamilton-Jacobi approach to potential functions outlined above has the advantage to clearly point out the mathematical regularity conditions needed to consider the problem well-posed from a not strictly local point of view.
Let us indulge a little on this subject.
On the one hand, the algorithm constructed in \cite{ay_amari-a_novel_approach_to_canonical_divergences_within_information_geometry}, as well as the one given above, heavily depends on the existence, uniqueness and differentiability properties of the exponential map $Exp_{\nabla}$ and of its inverse $Exp^{-1}_{\nabla}$.
These are strong assumptions that, from a global point of view, must be checked using a case-by-case analysis.
Furthermore, all the regularity requirements are relative to the affine connection $\nabla$, and thus the geometrical informations encoded in $g$ and $T$ seem to be uncorrelated, which is in contrast with the fact that the geometrical structure of a statistical manifold $(\mathcal{M}\,,g\,,T)$ considers $g$ and $T$ on the same footing.
On the other hand, the Hamilton-Jacobi approach to potential functions completely depends on a complete solution to the Hamilton-Jacobi problem for the Lagrangian $\lag$, which is a well-known problem.
The mathematical requirement for its well-posedness is the complete integrability (\cite{marmo_morandi_mukunda-a_geometrical_approach_to_the_hamilton-jacobi_form_of_dynamics_and_its_generalizations, arnold-mathematical_methods_of_classical_mechanics}) of the dynamical vector field associated with the Lagrangian $\lag$.
Consequently, once we write down the Lagrangian $\lag$, we are immediately able to pose the problem in a mathematically rigorous way, even if its explicit solution could be very hard to find.
Furthermore, the Lagrangian $\lag$ contains the informations about $g$ and $T$ together, which means that it contains all the informations on the geometrical structure of $(\mathcal{M}\,,g\,,T)$ as a whole. 
Having a Lagrangian, it would be possible to use the tools of symmetries and constants of the motion characteristic of Lagrangian mechanics to better understand the symmetry properties of the potential function $S_{\alpha}$.

\section{Conclusions and outlooks}

We have shown that a solution of the Hamilton-Jacobi problem for a Lagrangian $\lag_{\alpha}$ defined in terms of $g$ and $T$ is a potential function for the statistical manifold $\left(\mathcal{M}\,,g\,,T\right)$.
This dynamical perspective naturally leads to new questions, and, furthermore, paves the way to an interesting interchange of tools and methods between information geometry and the theory of dynamical systems.
For instance, the ``unfolding-reduction'' attitude towards dynamical systems clearly illustrated in \cite{carinena_ibort_marmo_morandi-geometry_from_dynamics_classical_and_quantum} could be a powerful technique in the search of potential functions.
Let us briefly comment on this point. 

Let us consider a two dimensional sphere embedded into $\mathbb{R}^3$ through the map $i_{S^2}\: : \mathbb{S}^2 \, \rightarrow \, \mathbb{R}^3$. A local expression of this map is given by

\be
\left\{\begin{array}{lcl} x^1 & = & \sin\theta \cos\phi \\ 
				   x^2 & = & \sin\theta \sin\phi \\ 
				   x^3 & = & \cos \theta
\end{array} \right.\,,
\ee
where $\theta \in \left] 0 , \pi \right[$ and $\phi \in \left] 0 , 2\pi \right[$. By means of this immersion it is possible to pull-back covariant tensors on $\mathbb{R}^3$ to $\mathbb{S}^2$. 

Let us consider the following statistical manifold: $\mathbb{R}^3$ equipped with the Euclidean metric $g= \delta_{jk}dx^j\otimes dx^k$ and the skewness tensor $T = dx^1\otimes dx^1\otimes dx^1 + dx^2\otimes dx^2 \otimes dx^2 + dx^3\otimes dx^3 \otimes dx^3$. According to the prescription outlined in this paper a canonical potential function is 

$$
S(x_{\mathrm{in}},x_{\mathrm{fin}})= \delta_{jk}(x_{\mathrm{fin}}^j-x_{\mathrm{in}}^j)(x_{\mathrm{fin}}^k-x_{\mathrm{in}}^k) + \dfrac{\alpha}{6} \left( (x^1_{\mathrm{fin}}-x^1_{\mathrm{in}})^3 + (x^2_{\mathrm{fin}}-x^2_{\mathrm{in}})^3 + (x^3_{\mathrm{fin}}-x^3_{\mathrm{in}})^3 \right)\,.
$$ 
By means of the previous immersion one can pull back this potential to $\mathbb{S}^2$ obtaining the following function

$$
S_{\mathbb{S}^2}((\theta_0, \phi_0),(\theta_1,\phi_1)) = \frac{1}{2}\left( \sin \theta_0 \sin \theta_1 \cos (\phi_1 - \phi_0) + \cos \theta_0 \cos \theta_1 \right) + 
$$
$$
+ \frac{\alpha}{6}\left( (\sin \theta_0 \cos \phi_0 - \sin \theta_1 \cos \phi_1)^3+(\sin \theta_0 \sin \phi_0 - \sin\theta_1 \sin \phi_1)^3+(\cos \theta_0 - \cos \theta_1)^3 \right)\,.
$$
A direct computation shows that this is a potential function on the submanifold $\mathbb{S}^2$ and it generates a metric tensor $g_{\mathbb{S}^2}$ and a skewness tensor $T_{\mathbb{S}^{2}}$ which coincide with the pull-back to $\mathbb{S}^2$ of the metric and skewness tensors on $\mathbb{R}^3$. Indeed

\be
g_{\mathbb{S}^2}= \mathrm{d}\theta\otimes\mathrm{d}\theta + \left( \sin \theta \right)^2\mathrm{d}\phi\otimes\mathrm{d}\phi\,,
\ee

$$
T_{\mathbb{S}^{2}}= - \left( \cos^3 \phi \cos^3\theta + \sin^3 \phi \cos^3 \theta -\sin^3 \theta \right)\mathrm{d}\theta\otimes\mathrm{d}\theta\otimes\mathrm{d}\theta \,-
$$
$$
- \, \cos^2 \theta \sin \theta \sin \phi \cos \phi (\cos \phi - \sin \phi)\mathrm{d}\theta\otimes\mathrm{d}\theta\otimes\mathrm{d}\phi \, + 
$$
$$
+ \,  \sin^2 \theta \cos \theta \sin \phi \cos \phi (\cos \phi + \sin \phi)\mathrm{d}\theta\otimes\mathrm{d}\phi\otimes\mathrm{d}\phi \, + 
$$
\be
+ \, \sin^3 \theta (\sin^3 \phi - \cos^3 \phi)\mathrm{d}\phi\otimes\mathrm{d}\phi\otimes\mathrm{d}\phi\,.
\ee

This simple example shows that in some cases it is possible to obtain a tensor which is no more constant, the metric tensor on the sphere, starting from an Euclidean space, and the potential on the Euclidean space induces a potential on the submanifold. However one could also invert this procedure. 
If one starts from a manifold with a non constant tensor it is possible to enlarge this manifold to a larger space equipped with a constant metric tensor: this is the meaning of the word ``unfolding'' in such a context.
These methods can be useful, for instance, in information geometry in relation with the description of curved exponential families, which are submanifolds of the statistical manifold of the exponential distribution. 
Another possible application is related to the Hamilton-Jacobi approach described in the previous sections. 
Indeed, we could enlarge the initial carrier space to a bigger space on which metric and skewness tensors are generated by a simple Lagrangian, simple in the sense of easily-solvable.
However, a better understanding of this situation can be achieved only by adopting an intrinsic language, and this is one of the possible developments we are working on.

Furthermore, the dynamical picture described in this contribution seems to suggest that the tangent bundle $T\mathcal{M}$ of the statistical manifold $\mathcal{M}$ plays an active role in the research of a contrast function $S$ for $(\mathcal{M}\,,g\,,T)$.
Consequently, it is natural to ask for a more clear interpretation of the tangent vectors to a probability distribution.

\vsp

The transition from the classical to the quantum setting is still to be fully worked out.
There are different aspects that need to be completely understood.
For instance, in the quantum setting the manifold $\mathcal{M}$ is the manifold of states of the system, hence, its points are no more probability distributions as in the classical case, they are probability amplitudes.
Indeed, denoting with $\psi(x)$  the wave function associated to a quantum state, it is well known that the square modulus $|\psi|^{2}\equiv \mathit{p}(x)$ of $\psi(x)$ is a genuine probability distribution.
Accordingly, the wave function can be written as $\psi(x)=\sqrt{\mathit{p}(x)}\,\mathrm{e}^{\mathrm{i} \alpha(x)}$, and thus a phase term $\mathrm{e}^{\mathrm{i} \alpha(x)}$ arises.
In \cite{facchi_kulkarni_manko_marmo_sudarshan_ventriglia-classical_and_quantum_fisher_information_in_the_geometrical_formulation_of_quantum_mechanics} it is shown that this phase term enters into the definition of the Fubini-Study metric $g$, as well as in the definition of the symplectic form $\omega$.
This, in turn, calls for a deeper understanding of the phase term in relation with the geometric structure of the manifold of quantum states.

\vsp  

Another question is related to some results known in Information Geometry and described, for instance in \cite{amari_nagaoka-methods_of_information_geometry} and \cite{amari-information_geometry_and_its_application}, i.e. that it is possible to use well-studied relative-entropies as contrast functions on a statistical manifold (examples are the Shannon relative entropy or the Tsallis q-relative entropy). 
Since we have interpreted contrast (potential) functions as solutions of a Hamilton-Jacobi problem, it is reasonable to ask whether relative entropies are generating functions of canonical transformations, and what such a transformation would do. 
This could then lead to a formulation of thermodynamics as a dynamical theory, entropy providing the action functional. 
Attempts in such direction have already been done. For instance Souriau in \cite{souriau-thermodynamique_et_geometrie} described thermodynamical evolutions in terms of symplectic scattering processes in a relativistic framework. 
However a deeper analysis in such a direction is necessary. 

\vsp

It would be interesting to apply the Hamilton-Jacobi procedure to formulate an inverse problem for well-known divergence functions (relative entropies) such as the Kullback-Leibler divergence.
Specifically, once a particular divergence function $S$ is fixed, the problem we have in mind is to find a suitable Lagrangian $\lag$ such that the associated Hamilton principal function $I$ is precisely the diverence function $S$ of the problem.
A preliminary step in this direction has been made in section \ref{sec: discussion} where it is shown that, if the Kullback-Leibler divergence $S_{KL}$ is the Hamilton principal function associated with a Lagrangian $\lag_{KL}$, then the Lagrangian $\lag_{\alpha}$ of equation \eqref{Lagrangian} differs from the third order approximation of $\lag_{KL}$ only by the constant factor $\left. \lag_{KL}\right|_{\mathbf{v}=0}$.

\vsp

Finally, let us note that the Hamilton-Jacobi approach to potential functions introduced here makes use only of the geometric informations contained in the metric tensor $g$ and the skewness tensor $T$, and it is independent on whether or not the manifold is a statistical manifold equipped with the Fisher-Rao metric.
This paves the way to a deeper analysis of the space of states of quantum mechanics where these structures are also present but do not necessarily coincide with the ``classical'' ones that appear in the context of statistical manifolds.

\section*{Acknowledgements}

We thank the anonymous referee for the valuable comments.
G.Marmo would like to acknowledge the support provided by the Banco de Santander-UCIIIM ”Chairs of Excellence” Programme 2016-2017.
J.M.P.P. is partly supported by the Spanish MINECO grant MTM2014-54692-P and QUITEMAD+, S2013/ICE-2801.

\addcontentsline{toc}{section}{References}
\bibliographystyle{unsrt}

\end{document}